\documentstyle[11pt]{article}
\newcommand{\ga}{$g_{\pi NN}(q^2)$}

\begin{document}

\begin{flushright}
       {\bf UK/94-02}  \\
 July 1994      \\
      hep-lat/9408007
\end{flushright}
\begin{center}

{\bf {\LARGE Nucleon Structure from Lattice QCD}}

\vspace{1cm}

{\bf    Keh-Fei Liu
 \footnote{ Invited talk at Energy Research Power Users
 Symposium, Rockville, MD, July 12, 1994}}\\ [0.5em]
 {\it  Dept. of Physics and Astronomy  \\
  Univ. of Kentucky, Lexington, KY 40506}

\end{center}


\begin{abstract}

We report results on the nucleon structure obtained from the lattice
quantum chromodynamics calculation. These include the axial,
electromagnetic, $\pi NN$, and scalar form factors. The
calculation is carried out at $\beta = 6$ on a $16^3 \times 24$
lattice with 24 quenched gauge configurations.  The chiral limit
results are extrapolated from several light quark cases. For the
disconnected insertion (sea-quark contribution), we used the
stochastic estimation with the $Z_2$ noise to calculate the diagonal
and off-diagonal traces of the inverse matrices with a size of $10^6
\times 10^6$. It is found that the $Z_2$ noise is the optimal choice
and its comparison with the Gaussian noise for our quark matrix is
given. For the sea-quark contribution, we report results on the
strange condensate in the nucleon and the $\pi N \sigma$
term.

\vskip\baselineskip

\end{abstract}



\section{Introduction}

\begin{quote}

The Understanding cannot See. \\
The Senses cannot Think. \\
By their union only can Knowledge be produced.

\hspace{10cm} --- {\bf {\em Immanuel Kant}}

\vspace{2ex}
Imagination is more important than Knowledge.

\hspace{10cm} --- {\bf {\em Albert Einstein}}

\end{quote}

The above quotes can be used to describe
the two well recognized approaches to scientific research ---
Experimental and Theoretical. Particularly true to the quotes is
the field of strong interaction where there
has been an intense interplay between theory and experiment.
The advent of quantum chromodynamics(QCD), a
non-abelian gauge relativistic quantum field theory of quarks and
gluons (the constituents of protons and neutrons), in the early
seventies has offered the most promising description of strong
interaction dynamics and internal structure of hadrons both in logical
consistency and in scope\cite{hooft}.  It is widely accepted that it
is the correct theory of strong interaction.  This acceptance is
based on a substantial amount of experimental data accumulated and
refined over the last two decades.  It is particularly so for the
inclusive scattering processes at large momentum transfer
where the perturbative QCD is applicable.  On the other hand,
confirmation of QCD as the fundamental theory of strong interaction
has been somewhat hampered by the lack of analytical calculations of
the low frequency modes due to the inherent non-perturbative nature of
the theory.

The invention of lattice-regularized QCD\cite{wilson} with Monte Carlo
methods\cite{creutz} holds the promise of circumventing the difficulty
of obtaining analytic results via numerical simulation.  In the
lattice-regularized gauge theory, the continuous space-time is
discretized by a finite lattice with a periodic boundary condition.
As a consequence, the infinite dynamic degrees of freedom associated
with the space-time is reduced to a finite number which makes
numerical analysis feasible.  The recent advancement in supercomputer
technology and the DOE Grand Challenge Award have made it possible to
carry out large scale Monte Carlo simulations to calculate the
hadronic structure and interaction in the framework of lattice QCD.
This allows us to compare the lattice results with the experimental
data in a qualitative and semi-quantitative manner which is very
valuable in understanding physics from first principles free from the
uncertainty of models.

In the present paper, we shall report results on the study of the
nucleon structure
which includes the electromagnetic, axial, pseudoscalar, and scalar
form factors and compare them with the known experiments.
{}From the pseudoscalar form factor, we extract the pion-
nucleon-nucleon form factor. We also report results on the
sea-quark contributions with disconnected current insertions.
This entails quark-loop calculations
involving diagonal and off-diagonal traces of the inverse
quark matrices with a dimension $10^6 \times 10^6$.
This is carried out with the stochastic estimation algorithm
with the $Z_2$ noise. We show that
the $Z_2$ noise is actually the optimal choice and a direct
comparison with the Gaussian noise is given. From this calculation,
we obtain the strange quark condensate in the nucleon.

As we shall see the above computer simulation has already been very
helpful
in bridging between the theory and experiment via {\it ab initio}
calculations and is doing a job better than the models, this third
branch of scientific research will only get more mature
when both the computer hardware and algorithm improve with time. To
put the numerical simulation in perspective, we maintain:

\begin{quote}

Imagination may not be relevant. \\
Knowledge cannot predict.  \\
By simulating Imagination and testing against Knowledge  \\
only can Reality be re-created.

\end{quote}

\section{Formalism and Numerical Steps}

The Euclidean path integral
formulation of quantum field theory is employed to calculate the
Green's function.  In this way, the quantum field theory becomes a
problem in classical statistical mechanics.  Physical observables are
then extracted from the statistical correlation functions of composite
operators built on the fundamental dynamical variables in quark and
gluon fields. In the following, we present a synopsis of
the formalism and  related numerical steps.

The correlation functions have the following generic form
\begin{equation} \label{corr1}
< O_1 ( U, \bar{\psi} , \psi ) O_2 ( U, \bar{\psi} ,
\psi ) ... >
= { 1 \over Z } \int [dU] [ d\psi ][ d \bar{\psi} ] e^{ -
S_{G} - S_{F} }
 \bar{\psi_{\alpha}}
\bar{\psi_{\beta}} ...  T^{\alpha \beta ...}_{ \alpha^{\prime }
\beta^{\prime} ...} (U) ...  \psi_{\alpha^{\prime} }
\psi_{\beta^{\prime} } ...,
\end{equation}
where $Z$ is the partition
function.  $S_G $ is the action of the gauge link variable U and
$S_F $ the fermion action which is  bilinear
in the quark field variable $\psi$ and can be written in the form $
S_F =\bar{\psi} M[U] \psi$.  Since both the fermion action $S_F$ and
the operator product $ O_1 O_2 ...$ are bilinear in $\bar{\psi}$ and
$\psi$, the fermion integral over the anti-commuting Grassmann
variables $\bar{\psi}$ and $ \psi$  can be done analytically to give
\begin{equation}  \label{corr2}
< O_1 ( U, \bar{\psi} , \psi ) O_2 ( U, \bar{\psi} ,
\psi ) ... >
= { 1 \over Z } \int [dU] e^{ - S_{G} }  \det
M[U] Tr [ M^{-1} [U] M^{-1} [U] M^{-1} [U] ... T] .
\end{equation}
The path integral in eq.(\ref{corr2})
is an ordinary multiple integral over
the group manifold, in this case $SU(3)$ group.  One might think of
approximating the integral by summing over a sufficiently dense set of
mesh points, say 4 per variable.  However , this  ``modest'' goal is
well beyond the capacity of the existing supercomputers: for a (small)
$10^4$ lattice, it would require a calculation on $ 4^{320,000} =
10^{192,659} $ points.  Instead, one may use the important sampling
technique to select a comparatively small subset of gauge
configurations $(U_1, U_2, ... , U_N)$ among the ``important'' ones,
such that the probability of occurrence of a given configuration $U_i$
in this ensemble approaches the desired distribution
\begin{equation} \label{glue}
[dU] e^{ - S_G } \det M[U],
\end{equation}
for $N \to \infty $.  Then
the quantum average are given(for sufficiently large $N$ ) by averages
taken over the sample: that is
\begin{equation} \label{sum}
< O_1 O_2 ... > = {
1 \over N} \sum^N_{i=1} Tr[ M^{-1}[U_i ] M^{-1}[U_i ] ... T ] \pm O( {
1 \over \sqrt{N}} ) .
\end{equation}

There are three numerical steps to lattice gauge calculations.  The
first step is to prepare an ensemble of gluon configurations $\{ U_i
\}$ in the vacuum.  This is done using the
Cabibbo-Marinari\cite{cabibbo} pseudo-heatbath algorithm, a Monte
Carlo method which generates sets of link matrices $\{ U_i \}$
according to the distribution in eq.(\ref{glue}).  In principle, the
probability distribution includes a factor of $\det (M[U])$
 which represents the effect of quark-antiquark polarization
in the vacuum.  In the present simulations, we omit this factor as an
approximation since it would require more computation than we can do.
We will discuss the validity of this approximation when we come to the
results of the calculations. Omitting this determinant is referred to
as the quenched approximation.

The second step is to calculate the quark propagator matrix $
M^{-1}[U]$ in the prepared gluon $(U)$ background.  This is done with
the conjugate-gradient algorithm\cite{hestenes}.  This is an iterative
scheme which is capable of giving the desired solution to arbitrary
accuracy.  We typically find a few hundred iterations to be
sufficient.

The last step is to assemble the quark propagator matrices $ M^{-1}
[U_i ] $ and the the matrix $ T [U_i ]$ defined in eqs. (\ref{corr1})
 and (\ref{corr2}) to evaluate the correlation function in
 eq.(\ref{sum}) from which
we extract the physical quantities with statistical analysis.

\section{Form Factors of Nucleon}
\subsection{Formalism}
Electron and neutrino scatterings off the nucleon
target are useful probes of the nucleon internal structure by
measuring its electromagnetic and axial form factors.
These form factors can be
simulated in the lattice gauge calculation and compared with
experiments directly and thereby serve as a test to the
Monte Carlo calculation and the theory as well.
The $\pi NN$ coupling form factor is a
fundamental quantity in the pion-nucleon and nucleon-nucleon dynamics.
Yet it is not experimentally accessible directly and is poorly known
theoretically. From the calculation of the pseudoscalar form factor,
we extract the $\pi NN$ form factor with pion pole dominance. In this
case, it is a prediction from the fundamental theory and
will have great help in settling the large uncertainty over the
shape and the cutoff of the $\pi NN$ form factor. Furthermore,
there have been sprouting interest in the sea-quark contributions to
the nucleon structure, for example, the $\pi N \sigma$ term, the
strange condensate in the nucleon and the flavor-singlet
axial charge $g_A^1$. The recent deep inelastic scattering experiment
obtained a very small $g_A^1$ which suggests that most of the spin
of the proton does not come from the quark spin which is a great
surprise and has been dubbed the ``proton spin crisis''. Since no
model we know has a handle on this problem, we expect the answer
to come from the lattice QCD calculation.

The nucleon can be obtained from the following two- and
three-point correlation functions \cite{dwl90,wdl92,ldd94}
\begin{equation} \label{2pt}
G_{NN} (t, {\bf p} ) = \sum_{ {\bf x} } e^{ - i {\bf
p \cdot x }} <0| N_{\alpha}(x) \bar{N}_{\alpha} (0) |0>.
\end{equation}
\begin{equation} \label{3pt}
G_{NjN} (t_1 , t_2, {\bf p}, {\bf q} ) = \sum_{ {\bf
x}_1 ,{\bf x}_2 } e^{  i {\bf q \cdot x_1 } - i {\bf p \cdot x_2}}
 \Gamma^{\alpha\beta} <0| N_{\beta}(x_2) j_{\mu} (x_1)
\bar{N}_{\alpha} (0) |0>,
\end{equation}
where  $N_{\alpha}$ represents the proton or
neutron interpolating field with the Dirac component $\alpha$,
$\Gamma$ is a $4 \times
4$ matrix in the Dirac space and $j_{\mu}$ represents
 various bilinear quark currents in the form of $\bar{\Psi}\Sigma
\Psi$. By inserting
complete sets of states and setting $t_2 >> t_1 >> a$, the lattice
spacing, we obtain the electromagnetic
form factors $G_E (q^2 )$ and $G_M (q^2 )$, axail form factor
$g_A(q^2)$, pseudoscalar form factor $g_P(q^2)$, and scalar form
factor $g_S(q^2)$  from
the ratios of the three point functions involving
corresponding currents to two point
functions. The results corresponding to the physical quark masses
are obtained from the extrapolation
from those with heavier quarks to the chiral limit where the pion
mass is zero.

\subsection{Numerical details}
Our results are based on 24
quenched (ignoring the determinant in eq. (\ref{glue}) gauge
configurations on a $ 16^3 \times 24$ lattice and were calculated
using the Monte Carlo Cabibbo-Marinari pseudo-heatbath
algorithm\cite{cabibbo}.  The $SU(3)$ fundamental Wilson
action was used with periodic boundary conditions and the coupling
constant was set at $\beta=6.0$.  The gauge field was thermalized for
5000 sweeps from a cold start and 24 configurations separated by at
least 1000 sweeps were saved.  For the quarks we use periodic boundary
conditions in the spatial directions and ``fixed'' time boundary
conditions, which corresponds to setting the
quark couplings across the time edge to zero.
The origin of all quark propagators was chosen to
be at lattice time site 5; the secondary zero momentum nucleon source
was fixed at time site 20.  We expect that these positions are
sufficiently far from the lattice time boundaries to avoid nonvacuum
contaminations.  We used the red-black pre-conditioned
conjugate-gradient algorithm
for the quark propagator. For our
convergence criterion we demanded that the absolute sum of the squares
of the quark propagators be less than $ 5 \times 10^{-5}$ over 5
iterations.  As a check of the nucleon secondary source, we verified
current conservation for $ t_2 > t_1 > 0 $ to $O( 10^{-4})$.

\subsection{Stochastic Estimation with $Z_2$ Noise}

Our present space-time
lattice with merely the size of $16^3 \times 24$ gives a quark
matrix of the dimension $10^6 \times 10^6$ including the
spin and color degrees of freedom. While it is durable to calculate
the quark propagator, i.e. $M^{-1}(x,0)$ for a point source S at 0
with a reasonably small quark mass (e.g. a fraction of the strange
quark mass) on today's supercomputers, the quark propagator
$M^{-1}(x,y)$ from any point to any point is certainly unattainable.
For calculations of the 2-point functions (eq. (\ref{2pt}) and
3-point functions (eq. (\ref{3pt}) with connected
insertions for flavor non-singlet currents,
one can get by with the help of translational symmetry and uses only
$M^{-1}(x,0)$.
But there are cases where one can not rely on such a help.
These include the calculations of quark loops which
are space-time or space integrations of the fermion propagators.
Examples of interest in QCD include the quark condensate and the
topological susceptibility with the fermion method,
flavor-singlet
meson masses which involve disconnected quark loops in the two-point
functions, notably the U(1) problem., and the $\pi N$ $\sigma$ term
and the proton spin problem which involve quark loop contributions
in the three-point functions.

Instead of waiting for the advent of more powerful hardware,
we have employed the stochastic algorithm with the $Z_2$ noise to
estimate the quark loops~\cite{dl94}.
Stochastic approach to estimating the inverse of
an $N \times N$ matrix M entails the introduction of an ensemble of
L column vectors $\eta \equiv {\eta^1,...,\eta^L}$ (each of
dimension $N \times 1$) with the properties of a white noise, i.e.
$\langle\eta_i\rangle = 0 ,
 \langle\eta_i \eta_j\rangle = \delta_{ij}, $
The expectation value of the
matrix element $M_{ij}^{-1}$ can be obtained by solving for $X_i$ in
the matrix equations $MX = \eta$ with the L noise vectors $\eta$ and
then take the ensemble average with the j-th entry of $\eta$
\begin{equation}
E[M_{ij}^{-1}] = \langle \eta_j X_i\rangle = \sum_k M_{ik}^{-1} \langle
 \eta_j \eta_k \rangle = M_{ij}^{-1}.
\end{equation}
which is the matrix element $M_{ij}^{-1}$ itself.
In fact, it has been shown recently
{}~\cite{bmt93} that the variance of a inverted matrix
element due to the stochastic estimation is composed of two parts

\begin{equation}  \label{variance}
Var[M_{ij}^{-1}] = \frac{1}{L} \{[M_{ij}^{-1}]^2 C_2^2 +
\sum_{k \neq j} [M_{ik}^{-1}]^2\}.
\end{equation}
Whereas the second part is independent of the kind of noise used,
the first part is proportional to the square of the
diagonal error
 $C_2 = \sqrt{\frac{1}{N}\sum_i(\langle \eta_i\eta_i\rangle -1)^2}$.
Since $Z_2$, or $Z_N$ for that matter,
has no diagonal error, i.e. $C_2 = 0$, it produces a {\bf minimum}
variance. Other noises will have larger variances
due to the non-vanishing $C_2$. For example, $C_2 = \sqrt{2/L}$ for
the Gaussian noise with large and independent configurations.

The question remains as to whether the $Z_2$ noise is superior than
the other noises on a practical footing with reasonable small L.
To answer this question,
we have tested the usefulness of the $Z_2$ noise against
the Gaussian noise for small noise configurations L ($L \leq 100$)
and smaller quark masses. Plotted in the left column in Fig. 1 are
the accumulated averages for the estimate of the real part of the
diagonal trace $Re Tr \sum_{\vec{x}} M^{-1}(x,x)/V$ as a function
of L for the $Z_2$ and Gaussian noises with Wilson hopping parameters
$\kappa = 0.148, 0.152$ and 0.154 for a gauge configuration.
With $\kappa_c = 0.1568$, $\kappa = 0.152$ corresponds to
the strange quark mass and $\kappa = 0.154$ corresponds to a
mass about half of that. The straight line is
the result from the stochastic estimate at L = 300. We see that
the estimate from the $Z_2$ noise approaches the value at L= 300
(assumed to be the asymptotic value) faster than that of the
Gaussian noise. The corresponding jackknife errors for different L
(the right column in Fig. 1)
show that the $Z_2$ noise is consistently better than the Gaussian
noise by a factor of two for all three $\kappa's$. This means that
to achieve the same level of accuracy for the diagonal trace, one
needs statistics for the Gaussian noise about 4 times as much as for
the $Z_2$ noise. We have also examined the
 near diagonal trace related to the point-split axial current.
The result is similar to that of the diagonal trace~\cite{dl94}.

We have employed the $Z_2$ noise to calculate the quark loops in the
presence of the nucleon with L = 200 to 300 only.
Some of the results will be reported in the
next section. Compared with the brute force approach of inverting
the whole quark matrix, we have saved the computer time by as much
as a factor of 6000!

It is worthwhile noting that the stochastic estimation is
particularly successful for the trace (denotes as $Re \overline{\Psi}
\Psi$ in Fig. 1). This is due to the
translational, color and spin symmetries. As a result, the error
is proportional to $1/\sqrt{N}$ where N is the dimension of the matrix.
With $N = 1.18 \times 10^6$ in our case and  the ratio
$  \sum_{k \neq 1} [M_{k1}^{-1}]^2 /[M_{11}^{-1}]^2 = 0.8$, we
predict the error to signal ratio to be $\sqrt{1.5} \times
10^{-3}$ from eq. (\ref{variance}) for L = 1. This
agrees well with the numerical calculation shown in Fig. 1.
Given this level of accuracy, it is feasible to apply the stochastic
method to the calculation of the determinant, the eigenvalues, and
the eigenvectors of the matrix M which might not be feasible with other
algorithms.

\section{Results}

We present our results in three groups. The first group includes the
electromagnetic and axial form factors where the experimental
results are available. We shall compare them with our calculation
directly. The second group consists of the $\pi NN$ form factor
deduced from the pseudoscalar form factor where is no
direct experimental result available. The third group involves
flavor-singlet quantities which require the quark loop calculation
with disconnected insertions like the $\pi N \sigma$ and the strange
condensate in the nucleon.

\subsection{Electromagnetic and axial form factors}

In Fig.~2, we plot the isovector axial form factor
extrapolated to the chiral
limit for the local current (L.C.) and the point-split current
(P-S.C.) in comparison with the experimental
result.  In doing so, we have used the calculated nucleon mass to set
the scale for the momentum ~\cite{ldd94}.  In
addition, we also show the calculated electric form factor $G_E$
extrapolated to the chiral limit
and the corresponding experimental results.  The experimental
$g_A(q^2)$ has been measured in neutrino-neutron scattering and pion
electroproduction.  The neutrino data gives a good fit in the dipole
form up to $q^{2} = 3 {\rm GeV}^2/c^2$~\cite{baker}, i.e.\ $g_{A} (q^2)
= g_{A} (0)/ (1 - q^2/M_{A}^2)^2$, with the axial vector coupling
constant $g_{A}(0) =1.254\pm 0.006$ and $M_{A} = 1.032\pm 0.036 {\rm
GeV}$ (world average).  The new data from Brookhaven E734
experiment~\cite{ahr88} gives a value $M_{A} = 1.09 \pm 0.03 \pm 0.02$
which is higher than the previous world average.  Our fit of the axial
form factor to the dipole form yields $M_{A} = 1.09 \pm 0.05 {\rm GeV}$
for the P-S.C.\ and $M_A = 1.06 \pm 0.04 {\rm GeV}$ for the L.C.\ which
are closer to the new experimental fit of the Brookhaven E734
data~\cite{ahr88}. Similarly, the fitted dipole mass for $G_E$ is
$M_{E} = 0.857 \pm 0.036 {\rm GeV}$ which is close to the experimental
dipole mass of $0.828\pm 0.006 {\rm GeV}$~\cite{sim80}.
Comparing to the experimental value $g_A = 1.254(6)$, we find that the
calculated $g_A =1.20(11)$ and 1.18(11) from the P-S.C.\ and L.C.\ are
$4\%$ and $6\%$ smaller respectively.

\subsection{$\pi NN$ form factor and the induced pseudoscalar
form factor}

The $\pi NN $ form factor $g_{\pi NN}(q^2)$ is a
fundamental quantity in the low-energy pion-nucleon
and nucleon-nucleon dynamics.
Many dynamical issues like the $\pi N$ elastic and
inelastic scattering, NN potential, three
-body force (triton and $He^3$ binding energies), pion photoproduction
and electroproduction all depend on it.  Similarly, the
pseudoscalar form factor is important to testing low-energy
theorems, chiral Ward identity and the understanding of the explicit
breaking of the chiral symmetry.
Yet, compared with the
electromagnetic form factors and the isovector axial form factor of
the nucleon shown in the last section,
the pseudoscalar form factor $g_P(q^2)$ and
the $\pi NN$ form factor $g_{\pi NN}(q^2)$ are poorly known either
experimentally or theoretically.
Notwithstanding decades of interest and numerous work,
the shape and slope of $g_{\pi NN}(q^2)$ remain illusive and
unsettling. Upon parametrizing \ga\,\, in the monopole form or the
dipole form, the cutoff mass can differ as much as a factor of 3 in
different models.

The $\pi NN$ form factor (shown in Fig. 3) is obtained from the
pseudoscalar form factor with pion pole dominance~\cite{ldd94a}.
We fitted it with both a monopole form and a dipole form.
We found that the monopole form with $\Lambda_{\pi NN} = 0.75 \pm
0.14 \rm{GeV}$ agrees with the Goldberger-Treiman relation at
$q^2 = 0$ and the dipole form does not.
Extrapolation of the monopole fit to $q^2 = m_{\pi}^2$ gives
$g_{\pi NN} = 12.7 \pm 2.4$ which agrees well with the
phenomenological value of $13.40 \pm 0.17$.
$g_{\pi NN}(0) = 12.2 \pm 2.3$ which agrees with the GT relation.

   This is a good example where the important physical quantity
is not directly accessible experimentally or reliably obtainable
from the models, lattice calculation can fill the void and give
a reliable prediction.

\subsection{$\bar{s}s$ in the nucleon and $\pi N \sigma$ term}

The matrix element $\langle N|\bar{s}s|N\rangle$
 and the $\pi N \sigma$ term requires
the quark loop calculation with disconnected insertions.

For the disconnected insertion of the current $ J(\vec{x}, \tau)
 = \overline{\Psi}(\vec{x}, \tau)\Gamma \Psi(\vec{x}, \tau) $,
 we calculate the ratio of the three- and two-point functions.

\begin{equation} \label{ratio}
\frac{\langle N(t) \sum_{\tau,
\vec{x}} J (\vec{x}, \tau) N^{\dagger} (0) \rangle}
{\langle N(t) N^{\dagger} (0) \rangle}  - \sum_{\tau, \vec{x}}
  \langle J(\vec{x}, \tau) \rangle
 \longrightarrow_{ t >> a}  const + t \langle N |\overline{\Psi}
 \Gamma \Psi |N \rangle_{dis}.
\end{equation}

Here  $N$ is the nucleon interpolation field and $\vec{x}$ is
summed over to obtain the forward matrix element. Hence, the matrix
element can be obtained as the slope from the above ratio. Since we
use the fixed boundary condition for the quark field in the time
direction, $\tau$ is summed to 4 steps away from the time-boundary
in both ends to avoid the boundary effect.

Plotted in Fig. 4 are the ratios defined in eq. (\ref{ratio}) for
quark masses ranging from the charm (Wilson $\kappa = 0.120$)
to strange ($\kappa = 0.154$). We see clearly that the slope becomes
larger when the quark mass becomes smaller. Extrapolating to the
chiral limit, we obtain  $\langle N|\bar{u}u + \bar{d}d |N\rangle
_{dis} = 5.36 \pm 1.04$.
This is very large, about twice of the connected insertion
$\langle N|\bar{u}u + \bar{d}d |N\rangle_{con} = 2.86 (6)$. If we
take the quark mass to be 5 MeV, the calculated
$\pi N \sigma$ term is then $41.1 \pm 5.2$ MeV. This is quite in
agreement with the 45 MeV obtained from $\pi N$ scattering and is also
consistent with a similar calculation on a $12^3 \times 20$ lattice
with the volume source~\cite{fko94}.

    We have also calculated $\langle N|\bar{s}s|N\rangle$.
 In this case, the quark mass in the
loop due to the current self-contraction corresponds to the
strange with $\kappa = 0.154$. The
valence quarks in the nucleon interpolating filed are then
extrapolated to the chiral limit. For the strange condensate, we
obtain $\langle N|\bar{s}s |N\rangle =  1.72 \pm 0.28$. If the
strange quark mass is taken to be 130 MeV, then the strange will
contribute $\sim 224 \pm 36$ MeV to the nucleon mass. This constitutes
an appreciable percentage of the total nucleon mass.

\section{Computational Aspects and Conclusion}
We have demonstrated in
this paper that it is feasible to perform {\it ab initio} calculations
of the hadronic structure directly from the
fundamental field theory---quantum chromodynamics without having to
rely on models.
Notwithstanding the fact that the present lattice size is
modest ( e.g. $ 16^3 \times 24$), the quark masses used are
heavy compared to the physical situation, the quenched
approximation is used, and the $Z_2$ noise algorithm saved us 6000
times of the computer time for the quark loop calculation,
present calculations still
took substantial supercomputer resources to perform. For example,
the project on the quark loop calculation took 2000 C-90 CPU hours
at 1 GFLOP/processor speed to accomplish. For the $16^3 \times 24$
lattice, the memory requirement is 20 Megawords.

In order to test QCD
as the fundamental theory and make predictions to compare with
experiments beyond doubt, we need to push the calculation to a much
larger lattice, a smaller quark mass and with the full description of
the dynamic fermions (non-quenched approximation).  Each of these
directions will demand one or more orders of magnitude in speed and
memory than the present calculation requires.

This work is partially supported by DOE
Grant DE-FG05-84ER40154. The author is
indebted to S.J. Dong, T. Draper, W. Wilcox, and
C.M. Wu who collaborated on the projects
presented here.  He is grateful to DOE for the generous Grand
Challenge Award which makes the calculations possible.

Figure Caption

\noindent

\noindent
Fig. 1 The accumulated averages of the real part of the diagonal trace
as estimated by the $Z_2$ and Gaussian noises for three quark masses
as functions of L, the number of noise vectors, are
plotted in the left column. The right column shows the corresponding
jackknife errors as functions of L.

\noindent
Fig. 2 The calculated axial form factor $g_A(q^2)$ and the proton
electric form factor $G_E(q^2)$  as a function of the momentum
transfer $-q^2$. The solid and the dash-dotted lines are fit to
two experimental sets of data with different dipole masses. The
dashed curve is the fit to the experimental $G_E(q^2)$ with a dipole
mass of 0.828 GeV.

\noindent
Fig. 3 The $\pi NN$ form factor $g_{\pi NN}(q^2)$ at the quark mass
which corresponds to the physical pion mass. The solid/dashed line
represents the monopole/dipole fit.

\noindent
Fig. 4 The ratio in eq. (\ref{ratio}) for the scalar charge as a
function of t for various quark masses. The slope is denoted by M
and the $\chi^2$ per degree of freedom is also given.
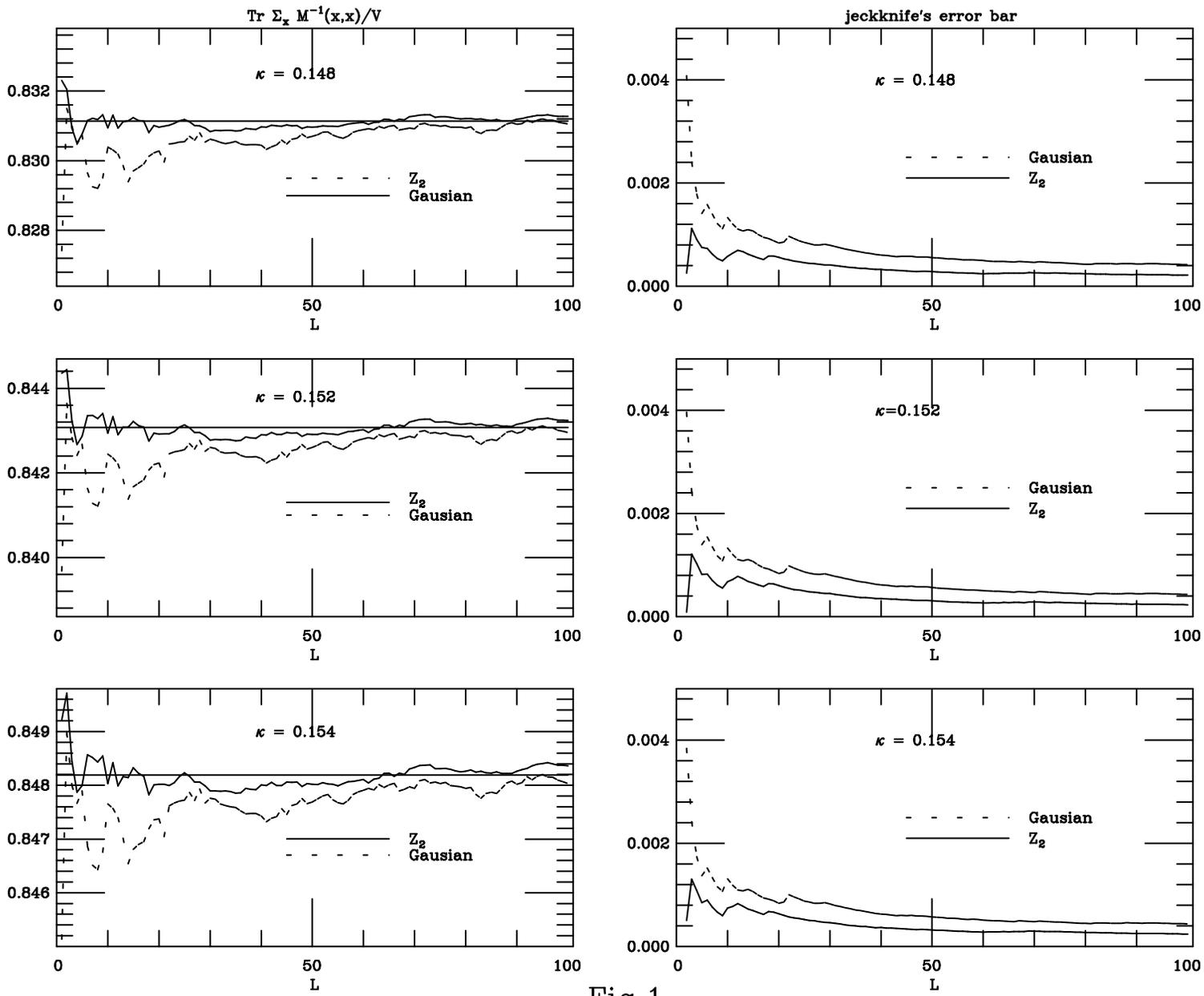
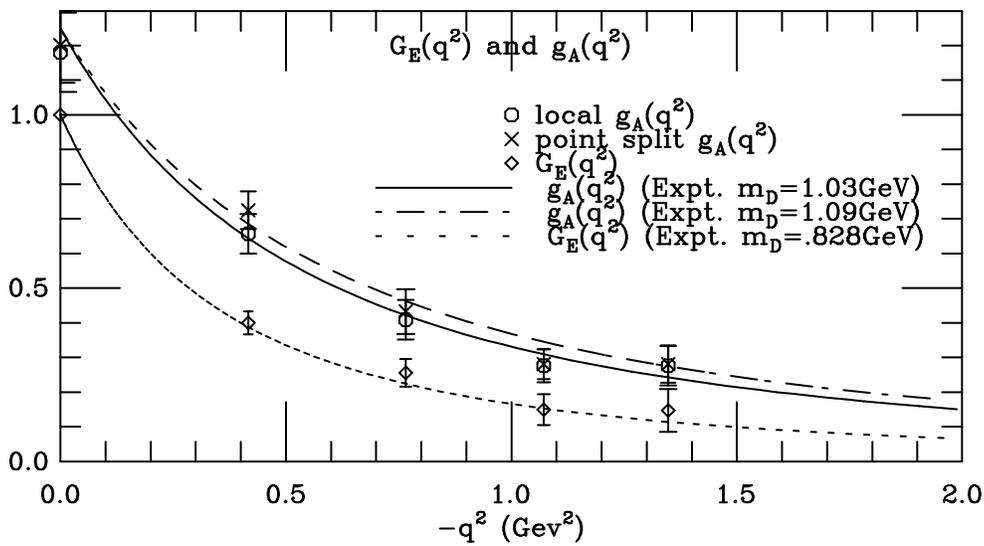
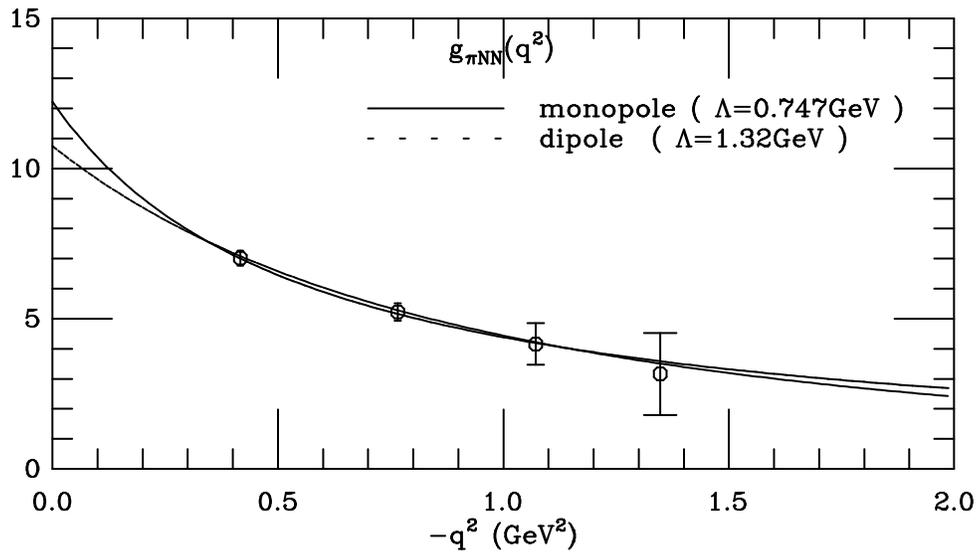
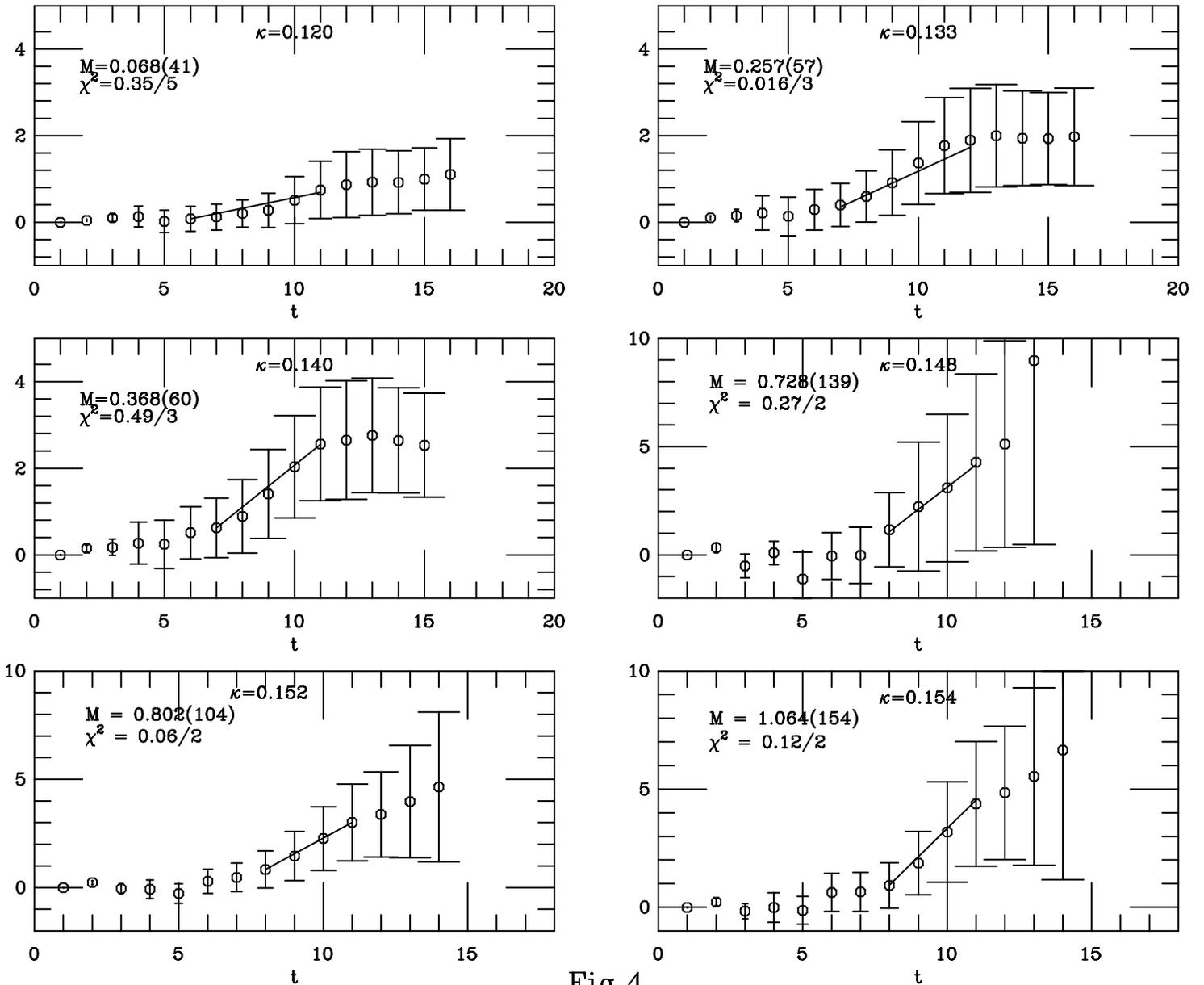
\end{document}